\documentclass[]{aa}

\usepackage{graphicx}
\usepackage{txfonts}

\begin{document} 

\title{On the local dark matter density}
\author{C. Moni Bidin\inst{1}
\and
R. Smith\inst{2}
\and
G. Carraro\inst{3}$^{,}$\inst{4}
\and
R. A. M\'endez\inst{5}\thanks{On leave at the European Southern Observatory,
  Casilla 19001, Santiago, Chile.}
\and
M. Moyano\inst{1}
}
\institute{
Instituto de Astronom\'ia, Universidad Catolica del Norte, Av. Angamos 0610, Antofagasta, Chile \\
\email{cmoni@ucn.cl}
\and
Departamento de Astronom\'ia, Universidad de Concepci\'on, Casilla 160-C, Concepci\'on, Chile
\and
European Southern Observatory, Alonso de Cordova 3107, Vitacura, Santiago, Chile
\and
Dipartimento di Fisica e Astronomia, Universit\'a di Padova, Vicolo Osservatorio~3, I-35122, Padova, Italia
\and
Universidad de Chile, Departamento de Astronom\'ia, Casilla 36-D, Santiago, Chile
}

\date{Received ; accepted ? }

\abstract
{In 2012, we applied a three-dimensional formulation to kinematic measurements of the Galactic thick disk and derived a surprisingly
low dark matter density at the solar position. This result was challenged by Bovy \& Tremaine (2012, ApJ, 756, 89), who claimed that
the observational data are consistent with the expected local dark matter density if a one-dimensional approach is adopted.}
{We aim at clarifying whether their work definitively explains our results, by analyzing the assumption at the bases of their
formulation and their claim that this returns a lower limit for the local dark matter density, which is accurate within 20\%.}
{We find that the validity of their formulation depends on the underlying mass distribution. We therefore analyze the predictions that
their hypothesis casts on the radial gradient of the azimuthal velocity $\partial_R\overline{V}$ and compare it with observational
data as a testbed for the validity of their formulation.}
{We find that their hypothesis requires too steep a profile of $\partial_R\overline{V}(Z)$, which is inconsistent with the
observational data both in the Milky Way and in external galaxies. As a consequence, their results are biased and largely overestimate
the mass density. Dynamical simulations also show that, contrary to their claims, low values of $\partial_R\overline{V}$ are compatible
with a Milky Way-like potential with radially constant circular velocity. We nevertheless confirm that, according to their criticism,
our assumption $\partial_R\overline{V}$=0 is only an approximation. If this hypothesis is released, and the available information about
$\partial_R\overline{V}$ in the thick disk is used, the resulting local dark matter density increases by a tiny amount, from $0\pm1$ to
$2\pm3$~mM$_\odot$~pc$^{-3}$, with an upper limit of $\sim$3.5~mM$_\odot$~pc$^{-3}$. Hence, this approximation has negligible influence
on our results.}
{Our analysis shows that their criticism is not a viable explanation for the inferred lack of dark matter at the solar position
detected by us. More studies are required to understand these unexpected results.}

\keywords{Galaxy: kinematics and dynamics --- dark matter --- Galaxy: structure}
   \maketitle


\section{Introduction}
\label{s_intro}
Dark matter (DM) plays a key role in many fields of modern astrophysics, and it is recognized as a fundamental component of the
Universe. Despite this general agreement, its density at the solar position ($\rho_{\odot,\mathrm{DM}}$) is still poorly constrained.
Even recent measurements are compatible within 1$\sigma$ with both a null local density \citep[e.g.,][]{Creze98,Holmberg00} and with a
value as high\footnote{The DM density will be given in the astronomical unit mM$_\odot$~pc$^{-3}$, where
1~mM$_\odot$~pc$^{-3}=10^{-3}$M$_\odot$~pc$^{-3}$=0.038~GeV~$c^{-2}$~cm$^{-3}$.} as 37~mM$_\odot$~pc$^{-3}$ \citep{Garbari12}. The
predictions of current models of spherical Galactic DM halo span the range $\rho_{\odot,\mathrm{DM}}$=5--13~mM$_\odot$~pc$^{-3}$
\citep[e.g.,][]{Olling01,Weber10}, but the density could be much lower in the case of a non-spherical or a more centrally concentrated
distribution \citep{Olling95,Einasto65,Einasto68}, or higher by up to a factor of two in the presence of dark substructures, such as a
dark matter disk or ring \citep{Kalberla03,Read08,Purcell09}.

\defcitealias{Moni12b}{M12b} \defcitealias{Moni12a}{M12a} \defcitealias{Dinescu11}{CD11}
Recently, \citet[][hereafter M12b]{Moni12b} have proposed a new formulation to measure the dynamical mass of the Galactic disk up to
large heights from the plane, by using the full three-dimensional kinematics and spatial distribution of a test stellar population. The
hypotheses at the basis of their calculations are not innovative, and they were (often implicitly) assumed in most of the previous
estimates. Applying this formulation to the kinematical measurements of \citet[][herafter M12a]{Moni12a}, \citet{Moni10,Moni12b} found
a surprising lack of DM at the solar position ($\rho_{\odot,\mathrm{DM}}=0\pm1$~mM$_\odot$~pc$^{-3}$), at variance with most of the
one-dimensional estimates even recently proposed \citep[e.g.,][]{Zhang12}. A strong effort followed their work to understand the
reliability of these unexpected results.

\citet{Sanders12} argues, by means of numerical simulations, that \citetalias{Moni12a} underestimated the gradients of the velocity
dispersions with Galactic height by up to a factor of three. His work impels further investigation to derive more reliable kinematical
data, but it has negligible effects on the results of \citetalias{Moni12b}. In fact, they obtained identical results by adopting
alternative data sets, where the kinematical quantities were measured with different methods. Moreover, the mass derived with their
three-dimensional formulation is largely insensitive to an isotropic change in the dispersions, because a steeper gradient of the
vertical component ($\sigma_\mathrm{W}$) induces an increase in the inferred mass that is compensated by the larger negative
contribution of the radial and azimuthal ones ($\sigma_\mathrm{U}$ and $\sigma_\mathrm{V}$, respectively).

If the vertical trend of $\sigma_\mathrm{W}$ is expressed by a linear relation
\begin{equation}
\sigma_\mathrm{W}=\sigma_\mathrm{W,0}+Z\cdot\frac{\partial\sigma_\mathrm{W}}{\partial Z},
\label{eq_siglinear}
\end{equation}
and we isolate the terms containing $\frac{\partial\sigma_\mathrm{W}}{\partial Z}=\partial_Z\sigma_\mathrm{W}$ in Equation~(14) of
\citetalias{Moni12b}, their contribution to the surface density is
\begin{equation}
\Sigma_\mathrm{W}(Z)\approx (\pi G)^{-1}\cdot\sigma_\mathrm{W,0}\cdot
\partial_Z\sigma_\mathrm{W}\cdot\left(\frac{Z}{h_\mathrm{Z,\rho}}-1\right),
\label{eq_siglinear}
\end{equation}
where we neglected quadratic terms in $\partial_Z\sigma_\mathrm{W}$, whose contribution is smaller by one order of magnitude. With the
parameter definition of \citetalias{Moni12b} and the kinematical values of \citetalias{Moni12a}, we obtain
$\Sigma_\mathrm{W}(\mathrm{4~kpc})\approx 20$~M$_\odot$~pc$^{-2}$. Thus, enhancing $\partial_Z\sigma_\mathrm{W}$ by a factor of three,
the DM density increases by $\Delta\rho_{\odot,\mathrm{DM}}\approx\frac{\Delta\Sigma_\mathrm{W}}{2Z}=5$~mM$_\odot$~pc$^{-2}$. However,
repeating the same exercise for the other two components, we find
($\Sigma_\mathrm{U}+\Sigma_\mathrm{V})\approx (-10-4)$~M$_\odot$~pc$^{-2}\approx -0.7\cdot\Sigma_\mathrm{W}$. As a consequence, the
increase of $\rho_{\odot,\mathrm{DM}}$ in response to an isotropic increase in the dispersion gradients is damped by 70\%, when
$\sigma_\mathrm{U}$ and $\sigma_\mathrm{V}$ are taken into account. Even enhancing the vertical gradients by a factor of three, the
derived local mass density increases by only 1.5~mM$_\odot$~pc$^{-3}$. A gradient of $\sigma_\mathrm{W}$ steeper than that of the other
components is required to obtain a significantly higher DM density. Figure~6 of \citet{Sanders12} also shows that the large
underestimate claimed by the author is introduced by the offset measurements at $Z\leq$2.5~kpc, while in the range $Z$=2.5--4~kpc, the
\citetalias{Moni12a} method underestimates the dispersions by only $\sim$5\%, and their gradient by $\sim$15\%, both compatible with
their quoted errors. This discontinuity at $Z$=2.5~kpc is not observed in the real data of \citetalias{Moni12a} (see their Figure~6),
where the gradients change by less than $0.5\sigma$ if the points at $Z\leq$2.5~kpc are excluded. This suggests that Sanders'
simulations did not reproduce the measurements at lower heights well and that they introduced a systematic offset not present in the
measurements of \citetalias{Moni12a}, whose gradient estimates should be accurate within 15\%.

\defcitealias{Bovy12}{BT12}
\citet[][hereafter BT12]{Bovy12} also argue that \citetalias{Moni12b} results are flawed. They claimed that the same data of
\citetalias{Moni12a} are fully consistent with the standard value $\rho_{\odot,\mathrm{DM}}=10$~mM$_\odot$~pc$^{-3}$, with a lower
limit of $\rho_{\odot,\mathrm{DM}}\geq 5$~mM$_\odot$~pc$^{-3}$, if a more classical one-dimensional approach is adopted. In this paper,
we analyze the hypotheses underlying their formulation in more detail, to verify that their criticism explains the results of
\citetalias{Moni12b} and that this is the correct solution to the puzzle. Following \citetalias{Moni12b} and \citetalias{Bovy12}, we
use the cylindrical Galactic coordinates ($R,\theta,Z$), where $R$ is the Galactocentric distance, $\theta$ is directed in the
direction of Galactic rotation, and $Z$ is positive toward the north Galactic pole. The respective velocity components are
($\dot{R},\dot{\theta},\dot{Z}$)=($U,V,W$).


\begin{figure}
\centering
\includegraphics[angle=-90,width=\hsize]{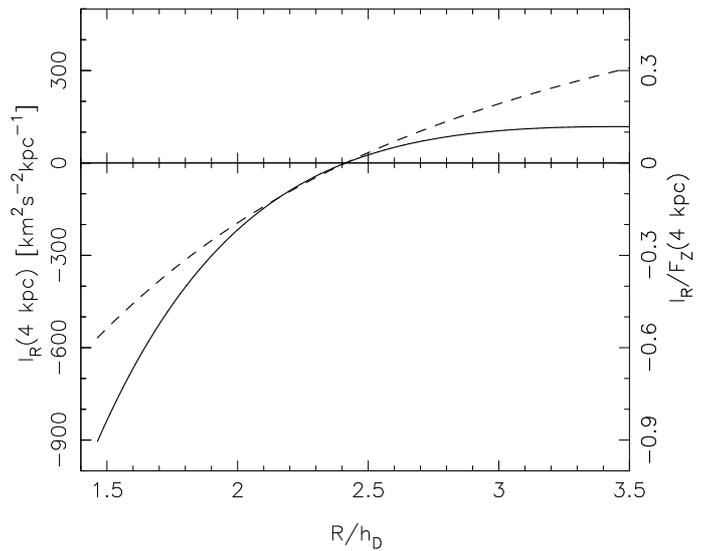}
\caption{Trend of $I_\mathrm{R}$ (solid line) and $\frac{I_\mathrm{R}}{F_\mathrm{Z}}$ (dashed line) at $Z$=4~kpc, for the \citet{Flynn96}
Galactic disk model, as a function of $R/h_\mathrm{D}$.}
\label{f_flynnIR}
\end{figure}

\section{Bovy \& Tremaine's assumptions}
\label{s_bovy}

Both \citetalias{Moni12b} and \citetalias{Bovy12} estimate the mass surface density $\Sigma (Z)$ within $\pm Z$~kpc of the Galactic plane by means
of the integrated Poisson equation in cylindrical coordinates:
\begin{equation}
2\pi G\Sigma(Z) =-I_\mathrm{R}(Z)-F_\mathrm{z}(Z),
\label{eq_Pois}
\end{equation}
with
\begin{equation}
I_\mathrm{R}(Z)=\int_{0}^{Z} \frac{1}{R}\frac{\partial (R F_\mathrm{R})}{\partial R}dz,
\label{eq_IR}
\end{equation}
where $G$ is the gravitational constant, and $F_\mathrm{R}$ and $F_\mathrm{Z}$ are the radial and vertical components of the force per unit mass,
respectively. If $F_\mathrm{R}$ in Equation~(\ref{eq_IR}) is expressed by means of the radial Jeans equation for a population in steady state,
\begin{equation}
F_\mathrm{R}=\frac{1}{\rho}\frac{\partial \left(\rho\overline{U^2}\right)}{\partial R}+
\frac{1}{\rho}\frac{\partial \left(\rho\overline{UW}\right)}{\partial Z}+\frac{\overline{U^2}-\overline{V^2}}{R},
\label{e_JeansR}
\end{equation}
the radial gradient of the mean azimuthal velocity $\frac{\partial\overline{V}(Z)}{\partial R}=\partial_R\overline{V}(Z)$ is
introduced in the formulation, because
\begin{equation}
\frac{\partial (R F_\mathrm{R})}{\partial R}=\frac{\partial}{\partial R}\left[
\frac{R}{\rho}\frac{\partial \left(\rho\overline{U^2}\right)}{\partial R}+\frac{R}{\rho}\frac{\partial \left(\rho\overline{UW}\right)}{\partial Z}+
\overline{U^2}\right]-\frac{\partial \overline{V^2}}{\partial R},
\label{e_dRFR}
\end{equation}
and $\overline{V^2}=\overline{V}^2+\sigma^2_\mathrm{V}$. This quantity is poorly known observationally, and \citetalias{Moni12b} assumed
$\partial_R\overline{V}(Z)=0$ at any $Z$. They referred to $\overline{V}$ as the ``rotational'' velocity and to $\overline{V}(R)$ as the
``rotation curve'' henceforth. This terminology is not new in the literature \citep[e.g.,][]{Yoachim2005}, but very unfortunate, because
the same terms are used more often for the {\it circular} velocity $V_\mathrm{c}$. Nevertheless, their Equation~(11) unambiguously stated
their hypothesis. \citetalias{Bovy12} exposed this source of confusion, and arguing against the \citetalias{Moni12b} assumption,
they removed the source of the problematic term, setting $I_\mathrm{R}(Z)=0$ at any $Z$. This reduced the three-dimensional formulation of
\citetalias{Moni12b} to a one-dimensional approach, similar to the previous works \citep[e.g.,][]{Kuijken89,Holmberg00}, but for the first time
extended beyond 1.1~kpc from the Galactic plane. They also claim that $I_\mathrm{R}(Z)<0$ and $\frac{I_\mathrm{R}}{F_\mathrm{z}}(Z)<0.2$.
The first inequality implies that their assumption $I_\mathrm{R}(Z)=0$ returns a lower limit of $\rho_{\odot,\mathrm{DM}}$, because
$I_\mathrm{R}$ enters with negative sign in the calculation of the surface mass density. In addition, if
$\frac{I_\mathrm{R}}{F_\mathrm{z}}(Z)<0.2$, the resulting estimate of $\Sigma(Z)$ is accurate within 20\% (see Equation~(\ref{eq_Pois})).
They claim that they only substitute the \citetalias{Moni12b} assumption with the observationally-proven fact
$\frac{\partial V_\mathrm{c}}{\partial R}(Z)=0$ \citep[e.g.,][]{Sofue09}, but all their calculations were actually based on the implicit
assumption that $-F_\mathrm{R} R=V_\mathrm{c}^2$ up to $Z$=4~kpc. However, this equation is strictly valid only on the plane. Assuming
$h_\mathrm{R}$=3.8~kpc and $h_\sigma$=3.5~kpc for the radial scale length of the thick disk mass density and velocity dispersions, respectively,
they derived $\rho_{\odot,\mathrm{DM}}=5.5$~mM$_\odot$~pc$^{-3}$, with a slope of $\Sigma(Z)$ compatible with
$\rho_{\odot,\mathrm{DM}}=7$~mM$_\odot$~pc$^{-3}$. They also argue that $h_\mathrm{R}$=2~kpc should be preferred, and they obtained
$\rho_{\odot,\mathrm{DM}}=8.5\pm1.5$~mM$_\odot$~pc$^{-3}$ in this case. In this section, we analyze in more detail the \citetalias{Bovy12}
assumption $I_\mathrm{R}(Z)=0$ that leads to these results.

\begin{figure}
\centering
\includegraphics[angle=-90,width=\hsize]{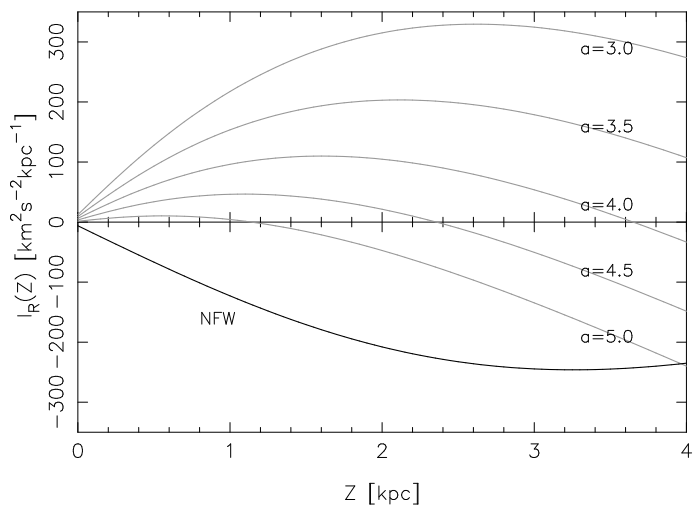}
\caption{Trend of $I_\mathrm{R}(Z)$ with $Z$ at $R$=8~kpc, for a Miyamoto-Nagai disk (gray curves) with $b$=0.3~kpc and
$a$=2.5, 3.5, 4.5~kpc, and a NFW halo model with $\rho_{\odot,\mathrm{DM}}=8$~mM$_\odot$~pc$^{-3}$ (black curve).}
\label{f_mn}
\end{figure}

\subsection{The universality of the $I_\mathrm{R}(Z)=0$ assumption}
\label{ss_models}

We first analyze the \citetalias{Bovy12} claim that $I_\mathrm{R}(Z)<0$ and $\frac{I_\mathrm{R}}{F_\mathrm{z}}(Z)<0.2$.
The sign of $I_\mathrm{R}(Z)$ and its weight in the calculation of $\Sigma(Z)$ depend critically on the mass density distribution.
\citetalias{Bovy12} prove the validity of their claims at $R=8$~kpc for an exponential disk model with scale length $h_\mathrm{R}$=3.4~kpc
and for a spherical NFW \citep{Navarro97} dark halo. However, alternative mass distributions lead to completely different
conclusions. For example, as shown in Fig.~\ref{f_flynnIR}, $I_\mathrm{R}(Z)$ is positive up to $Z$=4~kpc for the Galactic disk model of
\citet{Flynn96} at $R/h_\mathrm{D}>2.4$, where $h_\mathrm{D}$ is the radial scale length of the disk mass density. This condition
is most likely verified at the solar position, where $R\approx$8--8.5~kpc and $h_\mathrm{D}$=2--3~kpc \citep{Juric08,Bovy12b}.
Figure~\ref{f_flynnIR} also shows the trend of $\frac{I_\mathrm{R}}{F_\mathrm{z}}(Z)$, which exceeds 0.2 if $R/h_\mathrm{D}>3$, i.e. if
$h_\mathrm{D}\la$2.7~kpc.

In Fig.~\ref{f_mn} we show the trend of $I_\mathrm{R}(Z)$ at $R=8$~kpc for a family of Miyamoto-Nagai (MN) disk models
\begin{equation}
\Phi(R,Z)=\frac{-GM}{\sqrt{R^2+\left(a+\sqrt{Z^2+b^2}\right)^2}}
\label{eq_MN}
\end{equation}
\citep{Miyamoto75} with $b$=0.3~kpc and total mass $M=6\cdot10^{11}$M$_{\odot}$. The figure shows that $I_\mathrm{R}(Z)$ is positive up
to $Z$=3.5~kpc for the MN disk with $a$=4~kpc used by \citet{Qu11} and \citet{Bovy12b} to model the Galactic disk. More in general,
$I_\mathrm{R}(Z)>0$ up to $Z$=4~kpc for a MN disk with $a<$4~kpc. The
parameter $a$ must be increased to $a/R_0>$0.63 if we want $I_\mathrm{R}(Z)<0$ at $R=R_0$ in the whole range $Z$=0--4~kpc.

The sign of $I_\mathrm{R}(Z)$ and its incidence in Equation~(\ref{eq_Pois}) depend on the relative weight of the individual mass components,
due to the additive property of the potential. A spherical extended feature such as a DM halo gives a negative contribution, as
shown for example in Fig.~\ref{f_mn} for a NFW dark halo with $\rho_{\odot,\mathrm{DM}}$=8~mM$_\odot$~pc$^{-3}$. If the Galactic disk is
approximated by the aforementioned MN model with $a$=4, the presence of such a dark halo is required to have $I_\mathrm{R}(Z)<0$ up to
$Z$=4~kpc.

It must be considered that, when $I_\mathrm{R}(Z)$ is positive and neglected, the {\it total} mass is overestimated, leading to a larger
overestimate of $\rho_{\odot,\mathrm{DM}}$, because the spurious excess of visible mass is ascribed to a higher DM density. For example, if
$\rho_{\odot,\mathrm{DM}}$=5~mM$_\odot$~pc$^{-3}$, a mass overestimate of 20\% at $Z$=2.5~kpc would increase the derived
$\rho_{\odot,\mathrm{DM}}$ to 8.5~mM$_\odot$~pc$^{-3}$, thus overestimating the DM density by $\sim$70\%.

\begin{figure}
\centering
\includegraphics[angle=-90,width=\hsize]{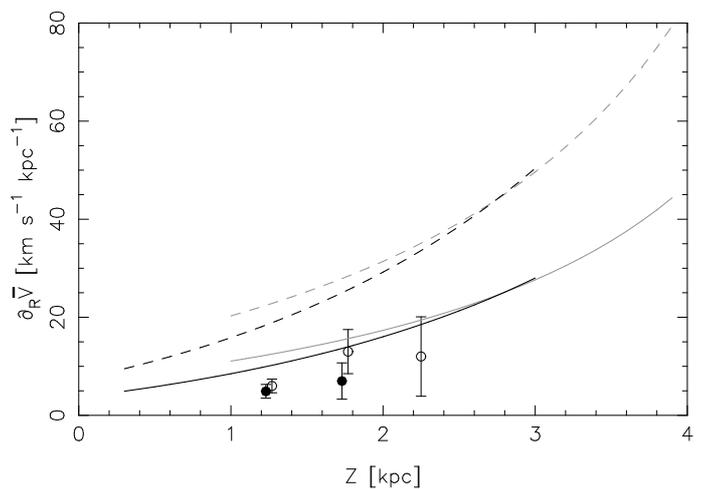}
\caption{Vertical trend of $\partial_R\overline{V}$ expected by the hypothesis $I_\mathrm{R}(Z)=0$, assuming the \citetalias{Moni12b} (full
curves) and \citetalias{Bovy12} (dashed curves) parameter set, and the kinematical results of \citet{Dinescu11} (black curves) and
\citetalias{Moni12a} (gray curves). The measurements of \citet{Dinescu11} are shown as empty dots, and their revision discussed
in Section~\ref{sss_Din} is indicated by full dots.}
\label{f_Din}
\end{figure}

In conclusion, our counterexamples show that $I_\mathrm{R}(Z)$ is not negative and negligible for any Galactic mass model. As a consequence,
the \citetalias{Bovy12} formulation is not generally valid. It is a good approximation if the underlying mass distribution is close
to the Galactic models implicitly assumed, but if it differs, the resulting estimate is not necessarily a lower limit to
$\rho_{\odot,\mathrm{DM}}$, nor is it accurate within 20\%. Such implicit assumptions should obviously be avoided in a formulation aimed at
measuring the local mass density itself.

\subsection{Tests of the validity of the $I_\mathrm{R}(Z)=0$ assumption}
\label{ss_tests}

Given the results of the previous section, we test the \citetalias{Bovy12} assumption here, to verify that their formulation is coincidentally
reliable in the specific case under study. The hypothesis $I_\mathrm{R}(Z)=0$ requires $\frac{\partial (R F_\mathrm{R})}{\partial R}=0$ at any $Z$.
When $F_\mathrm{R}$ is expressed by means of Equation~(\ref{e_JeansR}), this translates to:
\begin{equation}
2\overline{V}\partial_R \overline{V}=k_1\sigma^2_\mathrm{U}+\frac{\sigma^2_\mathrm{V}}{h_\sigma}
+\left(\frac{R}{h_\sigma}-1\right)\left(\frac{\overline{UW}}{h_\mathrm{Z}}-
\frac{\partial \overline{UW}}{\partial Z}\right),
\label{eq_dvdr}
\end{equation}
with
\begin{equation}
k_1=\frac{R}{h_\mathrm{R} h_{\sigma}}+\frac{R}{h^2_{\sigma}}-\frac{1}{h_\mathrm{R}}-\frac{2}{h_{\sigma}},
\label{eq_k1}
\end{equation}
where we assumed a mass density distribution described by a double exponential law
($\rho\propto\exp{(-R/h_\mathrm{R} -\vert Z \vert/h_\mathrm{Z}))}$ and an exponential radial decay of the velocity dispersions with scale length
$h_\sigma$. Both these assumptions were discussed, justified, and adopted by \citetalias{Bovy12} and \citetalias{Moni12b}.

Equation~(\ref{eq_dvdr}) must be satisfied if the \citetalias{Bovy12} hypothesis is correct. \citetalias{Bovy12} claim a good match between the
predictions on $\partial_R\overline{V}(Z)$ obtained from this equation and the observational results of \citet[][hereafter CD11]{Dinescu11}. Here
we study those predictions in more detail, re-analyze this comparison, and extend it to additional data sets. We assume
$h_\sigma$=3.8~kpc \citepalias{Moni12b,Bovy12} and a circular velocity at the solar position $V_\mathrm{c}=215\pm 30$~km~s$^{-1}$
\citep{Salucci2011}, which is indistinguishable from 220~km~s$^{-1}$ of \citetalias{Bovy12} for our purposes, but more compatible with the assumed
solar Galactocentric distance $R_\odot$=8~kpc \citep{Kerr86,Reid99}. Following the results of \citet{Sanders12} discussed in Sect.~\ref{s_intro},
we increase the velocity dispersions of \citetalias{Moni12a} by 5\%, and their vertical gradient by 15\%, although this correction has negligible
effects on the results.

In Fig.~\ref{f_Din}, we show the expectations for $\partial_R \overline{V}(Z)$ obtained by inserting in Equation~(\ref{eq_dvdr}) the kinematical
results of \citetalias{Moni12a} (their Equations (3) to (5), with $\overline{V}=190-(30\cdot Z)$~km~s$^{-1}$) and \citetalias{Dinescu11} (their
Table~1, with $\overline{V}=201-(25\cdot Z)$~km~s$^{-1}$), with $\overline{UW}(Z)$ from \citetalias{Moni12b} (compatible with both data sets, see
\citetalias{Moni12b}). The hypothesis $I_\mathrm{R}(Z)=0$ requires high values of $\partial_R \overline{V}(Z)$, and its steep increase with $Z$.
Assuming $h_\mathrm{R}$=3.8~kpc and $h_\mathrm{Z}$=0.9~kpc, it grows from $\approx$10~km~s$^{-1}$~kpc$^{-1}$ at $Z$=1~kpc to
$\approx$45~km~s$^{-1}$~kpc$^{-1}$ at $Z$=4~kpc. Assuming $h_\mathrm{R}$=2~kpc, $h_\sigma$=3.5~kpc, and $h_\mathrm{Z}$=0.7~kpc, as preferred by
\citetalias{Bovy12}, $\partial_R \overline{V}(Z)$ increases even more, up to $\approx$80~km~s$^{-1}$~kpc$^{-1}$ at $Z$=4~kpc. In other words, the
\citetalias{Bovy12} assumption requires that the thick disk rotation increases rapidly with distance from the Galactic center and much more
rapidly at larger Galactic heights. For any parameter set, the radial gradient of $\overline{V}$ is six to eight times higher at $Z$=4~kpc than
on the Galactic plane. Thus, while the thick disk rotation decreases with distance from the plane
by $\frac{\partial \overline{V}}{\partial Z}\sim -30$~km~s$^{-1}$~kpc$^{-1}$ at the solar position \citep[e.g.,][]{Majewski92,Chiba00,Girard06},
this vertical shear would disappear within 3~kpc of the Sun ($R<$11~kpc), where the stars at $Z$=4~kpc would be corotating with those on the plane.
This shear would be inverted farther out, with stars outside the plane rotating faster. This peculiar thick disk kinematics have never been
observed in edge-on external galaxies, where $\overline{V}(R)$ is rather flat outside the central regions at any height \citep{Kregel04}, and the
off-plane thick disk rotation is always slower than on the plane with no appreciable change in the radial gradient \citep{Yoachim2005,Yoachim2008}.

\begin{figure}
\centering
\includegraphics[angle=-90,width=\hsize]{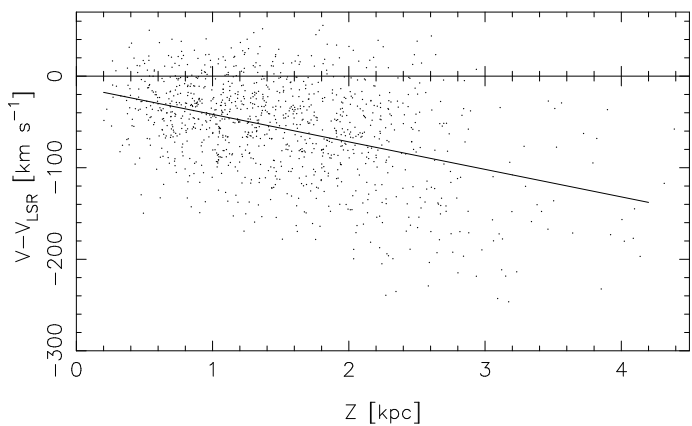}
\caption{Vertical trend of azimuthal velocity for our SDSS sample of thick disk stars. The linear fit to the data is indicated
by the solid line.}
\label{f_sdssdVdz}
\end{figure}

\subsubsection{Comparison with Casetti-Dinescu et al. (2011)}
\label{sss_Din}

The measurements of \citetalias{Dinescu11} for the Galactic thick disk are shown in Fig.~\ref{f_Din}. They confirm that the assumption
$\partial_R \overline{V}(Z)=0$ is only a rough approximation, because $\partial_R \overline{V}$ is in general not zero. However, the solution
predicted by the hypothesis $I_\mathrm{R}(Z)=0$, coupled with the \citetalias{Bovy12} preferred parameters, is clearly ruled out by the
observations. Even when assuming the parameter set criticized by \citetalias{Bovy12}, the inferred $\partial_R\overline{V}(Z)$ is systematically
higher than all the measurements, barely matching the two larger error bars. Moreover, the measurements of \citetalias{Dinescu11} are most likely
only upper limits of $\partial_R \overline{V}$. In fact, the radial profile of $\overline{V}$, similar to $V_\mathrm{c}$, is steeper in the inner
Galaxy and flattens at larger $R$ \citep[see, e.g., the theoretical radial profiles of][or our Fig.~\ref{f_vrsyn}]{Jalocha10}, and a linear fit in
the range $R$=6--9~kpc can easily overestimate $\partial_R\overline{V}(R_\odot)$. More reliable estimates can be derived by considering only the
stars within 1~kpc from the Sun ($R$=7--9~kpc). We thus obtain $\partial_R\overline{V}(Z=1.25$~kpc$)=4.9\pm 1.4$~km~s$^{-1}$~kpc$^{-1}$ and
$\partial_R\overline{V}(Z=1.75$~kpc$)=7\pm 4$~km~s$^{-1}$~kpc$^{-1}$, while the new fit at $Z$=2.25~kpc is affected by uncertainties that are too
large (close to 100\%) for a reliable re-estimate. As shown in Fig.~\ref{f_Din}, the expectations of the \citetalias{Bovy12} hypothesis are even
more discrepant when compared to these revised values. The comparison with the black curves is particularly relevant, because both the empirical
points and the expectations are obtained from the same data set.

\citetalias{Bovy12} performed a similar comparison with \citetalias{Dinescu11} data and claimed to prove the validity of their formulation. However,
their adopted input quantities differ from the results of \citetalias{Dinescu11} in many cases. For example, they assumed
$\sigma_\mathrm{U}$=60~km~s$^{-1}$ in the nearest bin, but Table~1 of \citetalias{Dinescu11} quotes 70.4~km~s$^{-1}$ (see also their Fig.12).
Similar problems can be found for $\sigma_\mathrm{V}$ and $\overline{V}$ in the same bin, all in the direction of decreasing the expected
$\partial_R \overline{V}(Z)$. Moreover, they fix $Z$ at the lower end of the \citetalias{Dinescu11} bins, thus decreasing the expectations further
by 2--3~km~s$^{-1}$~kpc$^{-1}$ in all the bins. They also never test their preferred set of parameters, which hugely offsets the expected
$\partial_R \overline{V}(Z)$ to higher values (Fig.~\ref{f_Din}).

\subsubsection{Comparison with SDSS data}
\label{sss_sdss}

\begin{figure}
\centering
\includegraphics[angle=-90,width=\hsize]{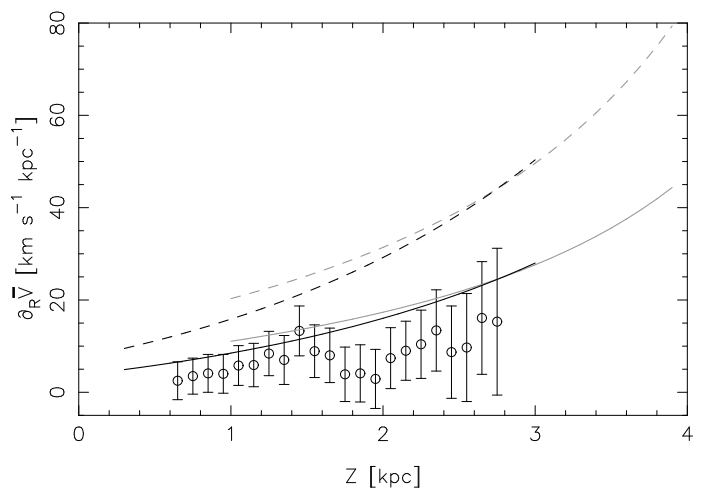}
\caption{Comparison of \citetalias{Bovy12} expectations for $\partial_R \overline{V}(Z)$ with results from SDSS data. The curves
are as in Fig.~\ref{f_Din}.}
\label{f_sdssdVdR}
\end{figure}

We collected from the Sloan Digital Sky Survey \citep[SDSS,][]{York00} DR7 database\footnote{http://www.sdss3.org/} the photometric and
spectroscopic data of stars with $r_0<$20.2, 0.48$<(g-r)_0<$0.55, S/N$>$15, $\log{g}>$4.2, 0.25$\leq$[$\alpha$/Fe]$\leq$0.3,
$-0.9\leq$[Fe/H]$\leq-0.5$, and error on metallicity and $\alpha$-elements abundance lower than 0.25~dex. These criteria were adopted to select
old, intermediate-metallicity G-type dwarf stars, as discussed in \citet{Carrell12} and \citet{Bovy12b}. Their distance was estimated
photometrically as in \citet{Ivezic08}. The Galactic coordinates, radial velocity, proper motion, and distance of each star were then transformed
into ($R,\theta,Z$) cylindrical coordinates and the respective ($U,V,W$) spatial velocities, and the errors on the former quantities were
propagated to derive the final uncertainties. Only the 1096 stars with $R$=7--9~kpc were considered in the analysis.

The azimuthal velocity of the resulting sample decreases with $Z$ by $-30\pm 2$~km~s$^{-1}$~kpc$^{-1}$, as shown in Fig.~\ref{f_sdssdVdz}. This
vertical shear is identical to what is found by \citetalias{Moni12a} (see \citetalias{Moni12a} for a comparison with previous studies). The
rotational properties of the two samples are therefore very similar, indicating that they probe the same Galactic stellar population. The SDSS
sample can therefore be used to study $\partial_R \overline{V}(Z)$ for the test population studied by \citetalias{Moni12b} and \citetalias{Bovy12}.

We fitted a linear relation to $\overline{V}(R)$ in $Z$ bins of width 0.5~kpc in steps of $\Delta Z$=0.1~kpc. A 2-$\sigma$ clipping algorithm was
adopted to remove outliers, which are mainly residual halo contaminators and bad measurements. The mean azimuthal velocity of these stars is
constant with $R$ in both cases. As a consequence, $\partial_R \overline{V}$ decreased at all heights if this clipping was not applied, while the
results were stable if the cut was stronger. The results are plotted in Fig.~\ref{f_sdssdVdR}, where they are compared to the expectations of
the \citetalias{Bovy12} assumption. The quantity $\partial_R \overline{V}(Z)$ increases slowly from $\sim$5~km~s$^{-1}$~kpc$^{-1}$ at $Z$=1~kpc to
$\sim$13~km~s$^{-1}$~kpc$^{-1}$ at $Z$=2.5--3~kpc. This trend matches the results of \citetalias{Dinescu11} extremely well. The SDSS data thus
confirm all the conclusions drawn in Sect.~\ref{sss_Din} and, in particular, confirm that the \citetalias{Bovy12} hypothesis largely overestimates
$\partial_R \overline{V}(Z)$ at any heights.

\subsubsection{Comparison with simulations}
\label{sss_sym}

\begin{figure}
\centering
\includegraphics[angle=-90,width=\hsize]{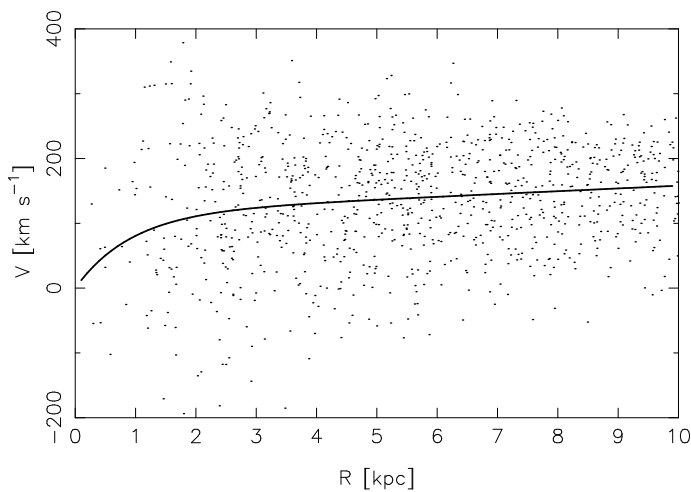}
\caption{Azimuthal velocity $V$ as a function of distance from the center, for the stars of our simulated sample in the bin $Z=$1.0--1.09~kpc. The
solid line indicates the linear fit in the range $R$=6--10~kpc.}
\label{f_vrsyn}
\end{figure}

\citetalias{Bovy12} claim that the \citetalias{Moni12b} hypothesis $\partial_R \overline{V}(Z)$=0 is inconsistent with a flat radial profile of
$V_\mathrm{c}$, while the steep increase in $\partial_R \overline{V}$ with $Z$ discussed in Sect.~\ref{ss_tests} is its natural consequence. We
tested this claim by means of orbit integration of a synthetic sample of stars. We adopted the Galactic potential model of \citet{Flynn96}, where
the disk is represented by the sum of three MN disks and the dark halo by a spherical logarithmic potential. The contribution of the Galactic bulge
and stellar halo are also included. The radial profile of $V_\mathrm{c}$ in this model Galaxy is flat outside the solar circle. The test particles
were spatially distributed as a double exponential disk with $h_\mathrm{Z}$=0.9~kpc and $h_\mathrm{R}$=3.5~kpc. Their initial kinematics was fixed
to match the trends of $\sigma_\mathrm{U}(Z)$, $\sigma_\mathrm{V}(Z)$, $\sigma_\mathrm{W}(Z)$, and $\overline{V}(Z)$ observed by \citetalias{Moni12a}
with the condition $\overline{U}=\overline{W}=0$ and an exponential radial decay of the dispersions with scale length $h_\sigma$=3.5~kpc. The system
was then left to relax in the potential, and it rapidly achieved its steady state within the first Gyr of integration, although the experiment was
stopped only after 5~Gyr. We then divided the sample in non-overlapping bins of 1000 stars with increasing $Z$, and we measured
$\partial_R \overline{V}$ in each bin fitting the function
\begin{equation}
V(R)=A\cdot\left[ 1-\exp{\left( -\frac{R}{B}\right) }\right]+C\cdot R,
\label{eq_Vsym}
\end{equation}
where $A$, $B$, and $C$ are the fit parameters. This functional form of the radial profile was chosen to account for the deviations from linearity at
lower $R$, as discussed in Sect.~\ref{sss_Din}. An example of these fits is shown in Fig.~\ref{f_vrsyn} for the 1000 stars with $Z$=2.0--2.1~kpc.

\begin{figure}
\centering
\includegraphics[angle=-90,width=\hsize]{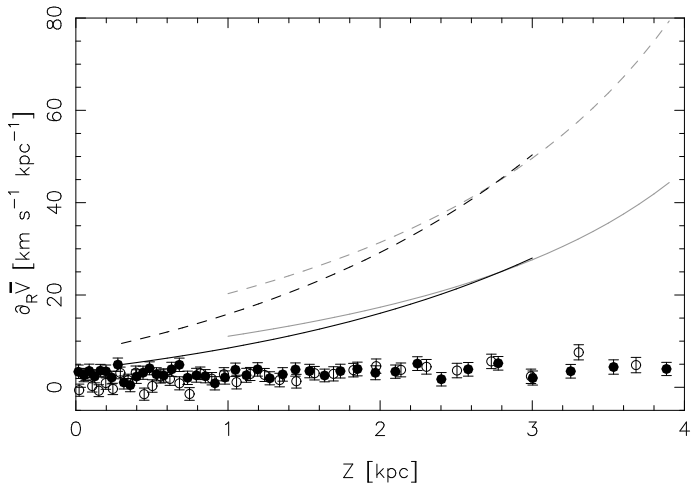}
\caption{Same as Fig.~\ref{f_Din}, but the expectations of \citetalias{Bovy12} are compared with the results of our symulations. The full dots
indicate the results when $\partial_R \overline{V}(Z)$=0 was assumed as initial condition, while the empty dots show the results when
$\partial_R \overline{V}(Z)$ is initially fixed as the black solid curve of this figure.}
\label{f_dVdRsym}
\end{figure}

The results of our experiment are shown in Fig.~\ref{f_dVdRsym}. We first imposed $\partial_R \overline{V}(Z)=0$ at any $Z$ as the initial condition,
and $\partial_R \overline{V}$ slightly increased during the integration. The system relaxed to tiny non-zero values with
$\partial_R \overline{V}(Z)\leq$5~km~s$^{-1}$~kpc$^{-1}$ up to $Z$=4~kpc. This result indicates that the \citetalias{Moni12b} hypothesis is wrong in
this specific Galactic model, because a flat radial gradient of $\overline{V}$ is not an equilibrium solution. In the second run, we imposed as
initial condition the lowest \citetalias{Bovy12} expectation, obtained from Equation~(\ref{eq_dvdr}) assuming the \citetalias{Moni12b} parameters and
the kinematical data of \citetalias{Dinescu11} (Figures~\ref{f_Din}, \ref{f_sdssdVdR}, and \ref{f_dVdRsym}). Very surprisingly, the system relaxed to
a $\partial_R \overline{V}(Z)$ profile identical to the first run, indicating that this is the real equilibrium configuration regardless of the
initial conditions. This solution is much lower than the curves expected by the $I_\mathrm{R}(Z)=0$ hypothesis, whose steep profile is therefore
unstable in this Galactic potential model. This test demonstrates that low values of $\partial_R \overline{V}(Z)$ are consistent with a Milky Way-like
galaxy, where $\frac{\partial V_\mathrm{c}}{\partial R}$=0 at the solar position, and a thick disk-like kinematics. The steep profile implied by the
\citetalias{Bovy12} formulation is therefore not a natural consequence of a radially constant circular velocity. It must be noted, however, that this
counterexample alone does not prove that $\partial_R \overline{V}(Z)$ {\it must} be small for any Galactic potential model and initial conditions.

\subsubsection{The validity of the $I_\mathrm{R}(Z)=0$ assumption}
\label{sss_conclusions}

We have shown that the observations, both in the Milky Way and in external galaxies, rule out the very peculiar thick disk kinematics required by the
hypothesis $I_\mathrm{R}(Z)=0$, which is at the basis of the \citetalias{Bovy12} formulation. Experiments of orbit integrations indicate that
it is not a requirement of a flat radial profile of $V_\mathrm{c}$, and it is even an unstable configuration in the Milky Way-like potential
adopted in the simulations.

The failure of the underlying assumption necessarily flaws the results of \citetalias{Bovy12}. It also invalidates the claim that their estimate is a
lower limit to $\rho_{\odot,\mathrm{DM}}$ accurate within 20\%. All evidence points to $I_\mathrm{R}(Z)>0$, because $\partial_R \overline{V}(Z)$ is
much lower than the values required to have $I_\mathrm{R}(Z)=0$. \citetalias{Bovy12} thus overestimate the mass density, because they
neglect a negative contribution in Equation~(\ref{eq_Pois}) or, equivalently, because they implicitly assume values of $\partial_R \overline{V}(Z)$
that are too high (see Figures~\ref{f_Din}, \ref{f_sdssdVdR}, and \ref{f_dVdRsym}), whose integral gives a positive contribution to $\Sigma(Z)$ (see
Equation~(\ref{eq_revised})). This bias is severe for their preferred solution (Figures~\ref{f_Din}, \ref{f_sdssdVdR}, and \ref{f_dVdRsym}), which
yields $\rho_{\odot,\mathrm{DM}}=8.5$~mM$_\odot$~pc$^{-3}$. However, this overestimate is present even when assuming $h_\mathrm{Z}$=0.9~kpc and
$h_\mathrm{R}$=3.8~kpc, and the corresponding \citetalias{Bovy12} solution for $\Sigma_\mathrm{>1.5kpc}(Z)$ (compatible with
$\rho_{\odot,\mathrm{DM}}\approx7$~mM$_\odot$~pc$^{-3}$) must also be biased.


\section{Moni Bidin et al. (2012) revisited}
\label{s_newest}

The results presented in Sect.~\ref{ss_tests} evidence that the \citetalias{Moni12b} hypothesis $\partial_R \overline{V}(Z)=0$ is only a rough
approximation, and a more precise estimate of the mass density can be derived if it is dropped. In this case, the \citetalias{Moni12b} expression
becomes
\begin{eqnarray}
\nonumber
2\pi G \Sigma(Z)=-\frac{k_1}{R}\cdot\int_{0}^{Z}\sigma_\mathrm{U}^{2}dz
-\frac{1}{R\cdot h_\sigma}\cdot\int_{0}^{Z}\sigma_\mathrm{V}^{2}dz+ \\
+k_3\cdot \overline{UW}+\frac{\sigma^{2}_\mathrm{W}}{h_\mathrm{Z}}-\frac{\partial \sigma^{2}_\mathrm{W}}{\partial Z}
+\frac{2}{R}\int_{0}^{Z}\overline{V}\left( \partial_R\overline{V}\right) dz,
\label{eq_revised}
\end{eqnarray}
where
\begin{equation}
k_3=\frac{1}{h_\mathrm{R}}-\frac{2}{R}+\frac{2}{h_\sigma}.
\label{eq_k3}
\end{equation}

The quantity $\partial_R\overline{V}(Z)$ cannot be estimated from \citetalias{Moni12a} data, but the \citetalias{Dinescu11} results can be used. In
fact, the two data sets probe the same stellar population (the intermediate-metallicity thick disk), as demonstrated by the fact that their
kinematical results are fully compatible. A small mismatch in the mean metallicity would only have tiny effects on the kinematics \citep{Bovy12c}. In
any case, our aim here is not a complete revision of the \citetalias{Moni12b} results, but to check to what extent these are biased by the assumption
$\partial_R \overline{V}(Z)=0$.

The studies of the Galactic disk measure the kinematical quantities only beyond a minimum Galactic height $Z_0$, because many observational
limitations prevent us from tracking their trend down to the Galactic plane. The estimate of $\Sigma(Z)$ thus requires the extrapolation of the
kinematics for $Z<Z_0$. Calculating the surface density of the mass enclosed between $Z_0$ and $Z$,
\begin{equation}
\Sigma_{>Z_0}(Z)=\int_{Z_0}^Z\rho(z)dz=\Sigma(Z)-\Sigma(Z_0),
\label{eq_relSigma}
\end{equation}
is more appropriate in this case. \citetalias{Moni12b} argue that their estimate of $\Sigma(Z)$ is not biased by the extension of the integration to
lower $Z$, and they obtained identical results analyzing both $\Sigma(Z)$ and $\Sigma_\mathrm{>1.5kpc}(Z)$. Nevertheless, the general reliability of
this extrapolation is not proven, and it is even less safe for Equation~(\ref{eq_revised}), which includes the previously neglected quantity
$\partial_R\overline{V}(Z)$. We therefore do not extend the calculation beyond the interval where we have information about it, and we study
$\Sigma_\mathrm{>1kpc}(Z)$ up to $Z$=2.5~kpc. \citetalias{Bovy12} also gave more importance to the increment of $\Sigma(Z)$ above $Z_0$ than to its
absolute value. Subtracting the expected visible mass from the result, as defined by \citetalias{Moni12b}, we derive the mean DM density in the range
$Z$=1--2.5~kpc, $\overline{\rho}_\mathrm{DM}$(1--2.5~kpc). It must be noted that the DM density derived from $\Sigma_{>Z_0}(Z)$ is nearly insensitive
to the assumed visible mass model when $Z_0\geq1$~kpc, because most of it lies below the volume under analysis. In fact, had different estimates
from the literature been assumed \citep[e.g.,][]{Holmberg00,Garbari11}, the $\overline{\rho}_\mathrm{DM}$(1--2.5~kpc) would vary by less than
0.2~mM$_\odot$~pc$^{-3}$.

\begin{figure}
\centering
\includegraphics[angle=-90,width=\hsize]{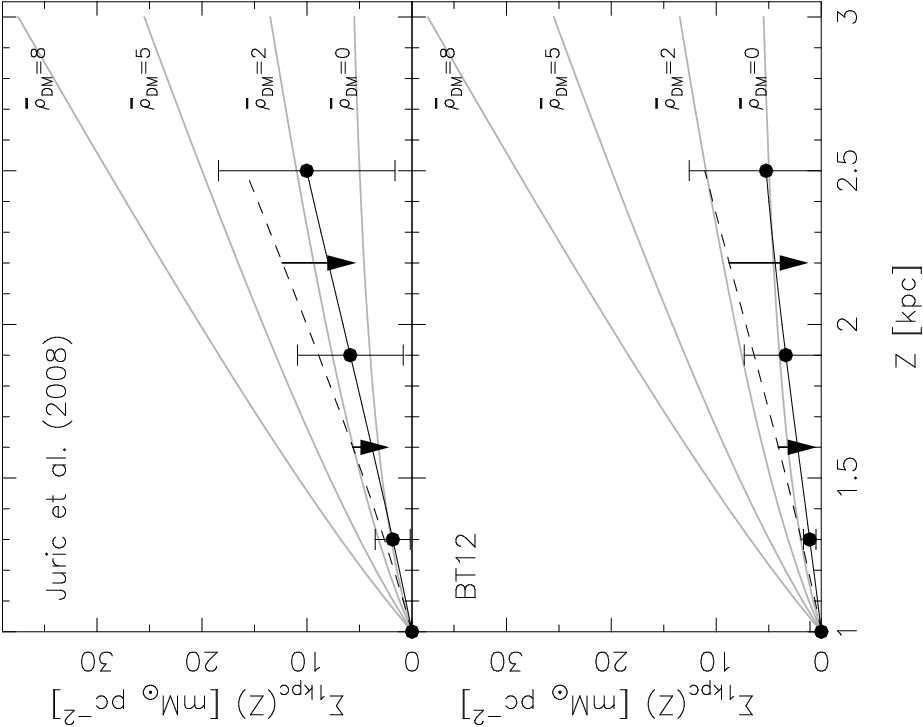}
\caption{Surface density of the mass at $Z>1$~kpc, calculated by inserting the kinematical results of \citetalias{Moni12a}, with the thick disk
parameters of \citet[][upper panel]{Juric08}, and \citetalias{Bovy12} (lower panel), in Equation~(\ref{eq_revised}). The dashed lines with arrows
indicate the upper limit derived from the estimates of \citetalias{Dinescu11} for $\partial_R\overline{V}(Z)$, while the black lines and dots with
error bars show the solution obtained from our re-analysis of this quantity. Curves of constant DM density (in mM$_\odot$~pc$^{-3}$) are overplotted
in gray.}
\label{f_resobs}
\end{figure}

Our results are shown in Fig.~\ref{f_resobs}, overplotted to a set of curves of constant DM density. We first estimated $\partial_R\overline{V}(Z)$ by
fitting the three measurements of \citetalias{Dinescu11} shown in Fig.~\ref{f_Din}. By inserting the result in Equation~(\ref{eq_revised}) along with
the other kinematical quantities from \citetalias{Moni12a} and the thick disk parameters preferred by \citetalias{Bovy12} ($h_\mathrm{R}$=2.0~kpc,
$h_\sigma$=3.5~kpc, and $h_\mathrm{Z}$=0.7~kpc), the curve of $\Sigma_\mathrm{>1kpc}(Z)$ returns
$\overline{\rho}_\mathrm{DM}=2\pm3$~mM$_\odot$~pc$^{-3}$. A higher DM density ($3.6\pm3.0$~mM$_\odot$~pc$^{-3}$) can be recovered assuming the
geometrical parameters $h_\mathrm{R}$=3.6~kpc and $h_\mathrm{Z}$=0.9~kpc from the extensive survey of \citet{Juric08}. Nevertheless, these results are
only upper limits because, as discussed in Section~\ref{sss_Din}, \citetalias{Dinescu11} probably overestimate $\partial_R\overline{V}(Z)$. If the
expression for $\partial_R\overline{V}(Z)$ is derived from the revised values of $\partial_R\overline{V}$ presented in Section~\ref{sss_Din}
(Fig.~\ref{f_Din}), we obtain $\overline{\rho}_\mathrm{DM}=0\pm2$~mM$_\odot$~pc$^{-3}$ when using the \citetalias{Bovy12} geometrical parameters, and
$\overline{\rho}_\mathrm{DM}=2\pm3$~mM$_\odot$~pc$^{-3}$ when adopting the values from \citet{Juric08}.

Similar results are found when $\sigma_\mathrm{U}(Z), \sigma_\mathrm{V}(Z)$, and $\sigma_\mathrm{W}(Z)$ are taken from \citetalias{Dinescu11}, and we
obtain $\overline{\rho}_\mathrm{DM}=4\pm5$~mM$_\odot$~pc$^{-3}$, with an upper limit of $\overline{\rho}_\mathrm{DM}<5.5\pm5.5$~mM$_\odot$~pc$^{-3}$.
These estimates are consistent with those obtained above, but they are poorly informative because of the large errors.


\section{Discussion and conclusions}
\label{s_conclusions}

We have shown that the validity of the \citetalias{Bovy12} hypothesis $I_\mathrm{R}(Z)=0$ depends on the underlying mass distribution. This means that
they implicitly constrain the mass distribution to measure the mass density in a given volume. Even the validity of their claim that their estimate is
a lower limit to $\rho_{\odot,\mathrm{DM}}$ accurate within 20\% depends on these constraints.

The assumption $I_\mathrm{R}(Z)=0$ predicts a very peculiar behavior for the Galactic thick disk rotation, ruled out by the observations of both the
Milky Way and external galaxies. The results of \citetalias{Bovy12} are flawed by this hypothesis. More specifically, they overestimate the mass
density, because their formulation implicitly assumes values of $\partial_R\overline{V}$ that are too large (see Equation~(\ref{eq_revised})). The
observations also indicate that the \citetalias{Moni12b} assumption $\partial_R\overline{V}(Z)=0$ is only a rough approximation, but we find that it
does not bias the results noticeably. In fact, when this assumption is dropped and $\partial_R\overline{V}(Z)$ is taken from literature measurements,
the resulting mean DM density at the solar position between $Z$=1 and 2.5~kpc is $\overline{\rho}_\mathrm{DM}=2\pm3$~mM$_\odot$~pc$^{-3}$, with an
upper limit of $\overline{\rho}_\mathrm{DM}<3.6\pm3.0$~mM$_\odot$~pc$^{-3}$. The thick disk parameters preferred by \citetalias{Bovy12} return much
lower values. These results agree well with those of \citetalias{Moni12b}, demonstrating that the incidence of $\partial_R\overline{V}(Z)$ on the
calculation is small. The resulting lack of DM is therefore not a systematic effect introduced by the aforementioned assumption. More investigation is
needed to understand the peculiar results found by \citetalias{Moni12b}.

The uncertainties quoted here and in \citetalias{Moni12b} result from rigorous propagation of errors on the input quantities, and they represent the
statistical uncertainties well. Possible systematic errors are not considered in the error budget, but \citetalias{Moni12b} analyzed most of their
hypotheses and found no relevant bias. Their uncertainties are small when compared to the literature, but it must be considered that their calculation
is extended in a volume that is four times larger than before. In fact, the extrapolation of the expected small DM contribution in the first kpc from
the Galactic plane (25\% of the total mass) is necessarily more uncertain than its estimate in a volume where it exceeds the quantity of visible mass.
For example, a 10\% uncertainty on the total mass and on the visible component propagates to a $\approx$40\% error on $\rho_{\odot,\mathrm{DM}}$ at
$Z<1$~kpc, but this reduces to 20\% and 10\% at $Z<$4~kpc, respectively. The volume analyzed here is smaller than \citetalias{Moni12b}, and the
resulting errors are larger by a factor of three. Our uncertainties are also partially enhanced by $\partial_R\overline{V}$, an additional source of
error not constrained very well by the data. As a result, our final errors are 30\% larger than those quoted by \citet{Zhang12}, although our volume
is comparable to theirs.

The results presented here are at variance with the expectations of a classical spherical DM halo with
$\rho_{\odot,\mathrm{DM}}>5$~mM$_\odot$~pc$^{-3}$, whose density decreases by only $\sim$5\% between $Z$=0 and 2~kpc. A deeper comprehension of the
power and limitations of the \citetalias{Moni12b} three-dimensional approach is required to fully understand this discrepancy. We nevertheless note
that, while a low density on the plane is compatible only with a highly prolate halo \citepalias[see][]{Moni12b}, a low mean density in the range
$Z$=1--2.5~kpc is compatible even with a very flat (oblate) distribution, if the bulk of the dark mass is found at $Z<1$~kpc. It is also important to
point out that the results obtained here are not directly comparable to previous estimates, because all similar measurements in the literature are
limited to $Z<1.2$~kpc.

All the literature measurements of the mass density at the solar position have adopted a one-dimensional approach, where the Galactic potential was
modeled to match the observational data \citep[but see][for an exception]{Korchagin03}. Both \citetalias{Bovy12} and \citetalias{Moni12b} make use of
a direct equation, which relates the mass density to the kinematical quantities. \citetalias{Bovy12} used the same kinematical data as
\citetalias{Moni12b}, and share most of the assumptions with them but, after reducing the three-dimensional formulation of \citetalias{Moni12b} to a
one-dimensional approximation, found results in agreement with both classical works \citep[e.g.][]{Kuijken89,Holmberg00} and more recent studies
\citep[e.g.][]{Siebert03,Zhang12}. Thus, they eventually prove that the discrepancy between the \citetalias{Moni12b} results and the literature is
most likely due neither to their use of a direct equation nor to a bias in the kinematical measurements. Their unexpected results most likely stem
from their innovative three-dimensional approach, and more investigation is needed to understand them.

\begin{acknowledgements}
R.A.M. acknowledges support from the Chilean Centro de Excelencia en Astrof\'isica y Tecnolog\'ias Afines (CATA) BASAL~PFB/06, and by Project
IC120009 ``Millennium Institute of Astrophysics (MAS)'' of the Iniciativa Cient\'{i}fica Milenio del Ministerio de Econom\'{i}a, Fomento y Turismo
de Chile. R.S. was financed by FONDECYT grant 3120135. Funding for SDSS-III has been provided by the Alfred P. Sloan Foundation, the Participating
Institutions, the National Science Foundation, and the U.S. Department of Energy Office of Science. R.A.M. acknowledges ESO/Chile for hosting him
during his sabbatical leave.
\end{acknowledgements}

\bibliographystyle{aa}
\bibliography{DMreply.bib}

\begin{thebibliography}{43}
\expandafter\ifx\csname natexlab\endcsname\relax\def\natexlab#1{#1}\fi

\bibitem[{{Bovy} {et~al.}(2012{\natexlab{a}}){Bovy}, {Rix}, {Hogg}, {Beers},
  {Lee}, \& {Zhang}}]{Bovy12c}
{Bovy}, J., {Rix}, H.-W., {Hogg}, D.~W., {et~al.} 2012{\natexlab{a}}, \apj,
  755, 115

\bibitem[{{Bovy} {et~al.}(2012{\natexlab{b}}){Bovy}, {Rix}, {Liu}, {Hogg},
  {Beers}, \& {Lee}}]{Bovy12b}
{Bovy}, J., {Rix}, H.-W., {Liu}, C., {et~al.} 2012{\natexlab{b}}, \apj, 753,
  148

\bibitem[{{Bovy} \& {Tremaine}(2012)}]{Bovy12}
{Bovy}, J. \& {Tremaine}, S. 2012, \apj, 756, 89

\bibitem[{{Carrell} {et~al.}(2012){Carrell}, {Chen}, \& {Zhao}}]{Carrell12}
{Carrell}, K., {Chen}, Y., \& {Zhao}, G. 2012, \aj, 144, 185

\bibitem[{{Casetti-Dinescu} {et~al.}(2011){Casetti-Dinescu}, {Girard},
  {Korchagin}, \& {van Altena}}]{Dinescu11}
{Casetti-Dinescu}, D.~I., {Girard}, T.~M., {Korchagin}, V.~I., \& {van Altena},
  W.~F. 2011, \apj, 728, 7

\bibitem[{{Chiba} \& {Beers}(2000)}]{Chiba00}
{Chiba}, M. \& {Beers}, T.~C. 2000, \aj, 119, 2843

\bibitem[{{Creze} {et~al.}(1998){Creze}, {Chereul}, {Bienayme}, \&
  {Pichon}}]{Creze98}
{Creze}, M., {Chereul}, E., {Bienayme}, O., \& {Pichon}, C. 1998, \aap, 329,
  920

\bibitem[{{Einasto}(1965)}]{Einasto65}
{Einasto}, J. 1965, Trudy Astrofizicheskogo Instituta Alma-Ata, 5, 87

\bibitem[{{Einasto}(1968)}]{Einasto68}
{Einasto}, J. 1968, Publications of the Tartu Astrofizica Observatory, 36, 414

\bibitem[{{Flynn} {et~al.}(1996){Flynn}, {Sommer-Larsen}, \&
  {Christensen}}]{Flynn96}
{Flynn}, C., {Sommer-Larsen}, J., \& {Christensen}, P.~R. 1996, \mnras, 281,
  1027

\bibitem[{{Garbari} {et~al.}(2012){Garbari}, {Liu}, {Read}, \&
  {Lake}}]{Garbari12}
{Garbari}, S., {Liu}, C., {Read}, J.~I., \& {Lake}, G. 2012, \mnras, 425, 1445

\bibitem[{{Garbari} {et~al.}(2011){Garbari}, {Read}, \& {Lake}}]{Garbari11}
{Garbari}, S., {Read}, J.~I., \& {Lake}, G. 2011, \mnras, 416, 2318

\bibitem[{{Girard} {et~al.}(2006){Girard}, {Korchagin}, {Casetti-Dinescu}, {van
  Altena}, {L{\'o}pez}, \& {Monet}}]{Girard06}
{Girard}, T.~M., {Korchagin}, V.~I., {Casetti-Dinescu}, D.~I., {et~al.} 2006,
  \aj, 132, 1768

\bibitem[{{Holmberg} \& {Flynn}(2000)}]{Holmberg00}
{Holmberg}, J. \& {Flynn}, C. 2000, \mnras, 313, 209

\bibitem[{{Ivezi{\'c}} {et~al.}(2008){Ivezi{\'c}}, {Sesar}, {Juri{\'c}},
  {Bond}, {Dalcanton}, {Rockosi}, {Yanny}, {Newberg}, {Beers}, {Allende
  Prieto}, {Wilhelm}, {Lee}, {Sivarani}, {Norris}, {Bailer-Jones}, {Re
  Fiorentin}, {Schlegel}, {Uomoto}, {Lupton}, {Knapp}, {Gunn}, {Covey},
  {Smith}, {Miknaitis}, {Doi}, {Tanaka}, {Fukugita}, {Kent}, {Finkbeiner},
  {Munn}, {Pier}, {Quinn}, {Hawley}, {Anderson}, {Kiuchi}, {Chen}, {Bushong},
  {Sohi}, {Haggard}, {Kimball}, {Barentine}, {Brewington}, {Harvanek},
  {Kleinman}, {Krzesinski}, {Long}, {Nitta}, {Snedden}, {Lee}, {Harris},
  {Brinkmann}, {Schneider}, \& {York}}]{Ivezic08}
{Ivezi{\'c}}, {\v Z}., {Sesar}, B., {Juri{\'c}}, M., {et~al.} 2008, \apj, 684,
  287

\bibitem[{{Ja{\l}ocha} {et~al.}(2010){Ja{\l}ocha}, {Bratek}, {Kutschera}, \&
  {Skindzier}}]{Jalocha10}
{Ja{\l}ocha}, J., {Bratek}, {\L}., {Kutschera}, M., \& {Skindzier}, P. 2010,
  \mnras, 407, 1689

\bibitem[{{Juri{\'c}} {et~al.}(2008){Juri{\'c}}, {Ivezi{\'c}}, {Brooks},
  {Lupton}, {Schlegel}, {Finkbeiner}, {Padmanabhan}, {Bond}, {Sesar},
  {Rockosi}, {Knapp}, {Gunn}, {Sumi}, {Schneider}, {Barentine}, {Brewington},
  {Brinkmann}, {Fukugita}, {Harvanek}, {Kleinman}, {Krzesinski}, {Long},
  {Neilsen}, {Nitta}, {Snedden}, \& {York}}]{Juric08}
{Juri{\'c}}, M., {Ivezi{\'c}}, {\v Z}., {Brooks}, A., {et~al.} 2008, \apj, 673,
  864

\bibitem[{{Kalberla}(2003)}]{Kalberla03}
{Kalberla}, P.~M.~W. 2003, \apj, 588, 805

\bibitem[{{Kerr} \& {Lynden-Bell}(1986)}]{Kerr86}
{Kerr}, F.~J. \& {Lynden-Bell}, D. 1986, \mnras, 221, 1023

\bibitem[{{Korchagin} {et~al.}(2003){Korchagin}, {Girard}, {Borkova},
  {Dinescu}, \& {van Altena}}]{Korchagin03}
{Korchagin}, V.~I., {Girard}, T.~M., {Borkova}, T.~V., {Dinescu}, D.~I., \&
  {van Altena}, W.~F. 2003, \aj, 126, 2896

\bibitem[{{Kregel} {et~al.}(2004){Kregel}, {van der Kruit}, \&
  {Freeman}}]{Kregel04}
{Kregel}, M., {van der Kruit}, P.~C., \& {Freeman}, K.~C. 2004, \mnras, 351,
  1247

\bibitem[{{Kuijken} \& {Gilmore}(1989)}]{Kuijken89}
{Kuijken}, K. \& {Gilmore}, G. 1989, \mnras, 239, 605

\bibitem[{{Majewski}(1992)}]{Majewski92}
{Majewski}, S.~R. 1992, \apjs, 78, 87

\bibitem[{{Miyamoto} \& {Nagai}(1975)}]{Miyamoto75}
{Miyamoto}, M. \& {Nagai}, R. 1975, \pasj, 27, 533

\bibitem[{{Moni Bidin} {et~al.}(2012{\natexlab{a}}){Moni Bidin}, {Carraro}, \&
  {M{\'e}ndez}}]{Moni12a}
{Moni Bidin}, C., {Carraro}, G., \& {M{\'e}ndez}, R.~A. 2012{\natexlab{a}},
  \apj, 747, 101

\bibitem[{{Moni Bidin} {et~al.}(2012{\natexlab{b}}){Moni Bidin}, {Carraro},
  {M{\'e}ndez}, \& {Smith}}]{Moni12b}
{Moni Bidin}, C., {Carraro}, G., {M{\'e}ndez}, R.~A., \& {Smith}, R.
  2012{\natexlab{b}}, \apj, 751, 30

\bibitem[{{Moni Bidin} {et~al.}(2010){Moni Bidin}, {Carraro}, {M{\'e}ndez}, \&
  {van Altena}}]{Moni10}
{Moni Bidin}, C., {Carraro}, G., {M{\'e}ndez}, R.~A., \& {van Altena}, W.~F.
  2010, \apjl, 724, L122

\bibitem[{{Navarro} {et~al.}(1997){Navarro}, {Frenk}, \& {White}}]{Navarro97}
{Navarro}, J.~F., {Frenk}, C.~S., \& {White}, S.~D.~M. 1997, \apj, 490, 493

\bibitem[{{Olling}(1995)}]{Olling95}
{Olling}, R.~P. 1995, \aj, 110, 591

\bibitem[{{Olling} \& {Merrifield}(2001)}]{Olling01}
{Olling}, R.~P. \& {Merrifield}, M.~R. 2001, \mnras, 326, 164

\bibitem[{{Purcell} {et~al.}(2009){Purcell}, {Bullock}, \&
  {Kaplinghat}}]{Purcell09}
{Purcell}, C.~W., {Bullock}, J.~S., \& {Kaplinghat}, M. 2009, \apj, 703, 2275

\bibitem[{{Qu} {et~al.}(2011){Qu}, {Di Matteo}, {Lehnert}, \& {van
  Driel}}]{Qu11}
{Qu}, Y., {Di Matteo}, P., {Lehnert}, M.~D., \& {van Driel}, W. 2011, \aap,
  530, A10

\bibitem[{{Read} {et~al.}(2008){Read}, {Lake}, {Agertz}, \&
  {Debattista}}]{Read08}
{Read}, J.~I., {Lake}, G., {Agertz}, O., \& {Debattista}, V.~P. 2008, \mnras,
  389, 1041

\bibitem[{{Reid} {et~al.}(1999){Reid}, {Readhead}, {Vermeulen}, \&
  {Treuhaft}}]{Reid99}
{Reid}, M.~J., {Readhead}, A.~C.~S., {Vermeulen}, R.~C., \& {Treuhaft}, R.~N.
  1999, \apj, 524, 816

\bibitem[{{Salucci} {et~al.}(2010){Salucci}, {Nesti}, {Gentile}, \& {Frigerio
  Martins}}]{Salucci2011}
{Salucci}, P., {Nesti}, F., {Gentile}, G., \& {Frigerio Martins}, C. 2010,
  \aap, 523, A83

\bibitem[{{Sanders}(2012)}]{Sanders12}
{Sanders}, J. 2012, \mnras, 425, 2228

\bibitem[{{Siebert} {et~al.}(2003){Siebert}, {Bienaym{\'e}}, \&
  {Soubiran}}]{Siebert03}
{Siebert}, A., {Bienaym{\'e}}, O., \& {Soubiran}, C. 2003, \aap, 399, 531

\bibitem[{{Sofue} {et~al.}(2009){Sofue}, {Honma}, \& {Omodaka}}]{Sofue09}
{Sofue}, Y., {Honma}, M., \& {Omodaka}, T. 2009, \pasj, 61, 227

\bibitem[{{Weber} \& {de Boer}(2010)}]{Weber10}
{Weber}, M. \& {de Boer}, W. 2010, \aap, 509, A25

\bibitem[{{Yoachim} \& {Dalcanton}(2005)}]{Yoachim2005}
{Yoachim}, P. \& {Dalcanton}, J.~J. 2005, \apj, 624, 701

\bibitem[{{Yoachim} \& {Dalcanton}(2008)}]{Yoachim2008}
{Yoachim}, P. \& {Dalcanton}, J.~J. 2008, \apj, 682, 1004

\bibitem[{{York} {et~al.}(2000){York}, {Adelman}, {Anderson}, {Anderson},
  {Annis}, {Bahcall}, {Bakken}, {Barkhouser}, {Bastian}, {Berman}, {Boroski},
  {Bracker}, {Briegel}, {Briggs}, {Brinkmann}, {Brunner}, {Burles}, {Carey},
  {Carr}, {Castander}, {Chen}, {Colestock}, {Connolly}, {Crocker}, {Csabai},
  {Czarapata}, {Davis}, {Doi}, {Dombeck}, {Eisenstein}, {Ellman}, {Elms},
  {Evans}, {Fan}, {Federwitz}, {Fiscelli}, {Friedman}, {Frieman}, {Fukugita},
  {Gillespie}, {Gunn}, {Gurbani}, {de Haas}, {Haldeman}, {Harris}, {Hayes},
  {Heckman}, {Hennessy}, {Hindsley}, {Holm}, {Holmgren}, {Huang}, {Hull},
  {Husby}, {Ichikawa}, {Ichikawa}, {Ivezi{\'c}}, {Kent}, {Kim}, {Kinney},
  {Klaene}, {Kleinman}, {Kleinman}, {Knapp}, {Korienek}, {Kron}, {Kunszt},
  {Lamb}, {Lee}, {Leger}, {Limmongkol}, {Lindenmeyer}, {Long}, {Loomis},
  {Loveday}, {Lucinio}, {Lupton}, {MacKinnon}, {Mannery}, {Mantsch}, {Margon},
  {McGehee}, {McKay}, {Meiksin}, {Merelli}, {Monet}, {Munn}, {Narayanan},
  {Nash}, {Neilsen}, {Neswold}, {Newberg}, {Nichol}, {Nicinski}, {Nonino},
  {Okada}, {Okamura}, {Ostriker}, {Owen}, {Pauls}, {Peoples}, {Peterson},
  {Petravick}, {Pier}, {Pope}, {Pordes}, {Prosapio}, {Rechenmacher}, {Quinn},
  {Richards}, {Richmond}, {Rivetta}, {Rockosi}, {Ruthmansdorfer}, {Sandford},
  {Schlegel}, {Schneider}, {Sekiguchi}, {Sergey}, {Shimasaku}, {Siegmund},
  {Smee}, {Smith}, {Snedden}, {Stone}, {Stoughton}, {Strauss}, {Stubbs},
  {SubbaRao}, {Szalay}, {Szapudi}, {Szokoly}, {Thakar}, {Tremonti}, {Tucker},
  {Uomoto}, {Vanden Berk}, {Vogeley}, {Waddell}, {Wang}, {Watanabe},
  {Weinberg}, {Yanny}, {Yasuda}, \& {SDSS Collaboration}}]{York00}
{York}, D.~G., {Adelman}, J., {Anderson}, Jr., J.~E., {et~al.} 2000, \aj, 120,
  1579

\bibitem[{{Zhang} {et~al.}(2013){Zhang}, {Rix}, {van de Ven}, {Bovy}, {Liu}, \&
  {Zhao}}]{Zhang12}
{Zhang}, L., {Rix}, H.-W., {van de Ven}, G., {et~al.} 2013, \apj, 772, 108

\end{thebibliography}

\end{document}